# Infrared spectroscopy of cataclysmic variables - II. Intermediate polars

V. S. Dhillon[1], T. R. Marsh[2], S. R. Duck[3]★ and S. R. Rosen[4]

[1] *Royal Greenwich Observatory, Madingley Road, Cambridge CB3 0EZ*
[2] *Astrophysics, Physics Department, The University, Southampton SO9 5NH*
[3] *Astrophysics, Nuclear and Astrophysics Laboratory, Keble Road, Oxford OX1 3RH*
[4] *Astronomy Department, University of Leicester, University Road, Leicester LE1 7RH*

**ABSTRACT**
We present infrared (0.97–2.45 micron) spectra of the intermediate polars PQ Gem (RE0751+14), BG CMi and EX Hya. The spectra show strong Paschen, Brackett and He I emission lines from the accretion disc/stream. The infrared continua of PQ Gem and BG CMi can be represented by blackbodies of temperatures 4500 K and 5100 K, respectively, or by power-laws of the form $f_\nu \propto \nu^{0.6, 0.9}$, respectively, and show no evidence of secondary star features. The continuum of EX Hya is dominated by water bands from the red-dwarf secondary star, which has a spectral type of ∼M3. Despite showing circular polarization, PQ Gem and BG CMi show no evidence for cyclotron humps and hence we are unable to measure their magnetic field strengths; any cyclotron emission present must contribute less than ∼3% of the infrared continuum flux.

## 1 INTRODUCTION

Cataclysmic variables (CVs) are interacting binary systems in which a white dwarf primary accretes material from a red dwarf secondary. The path that the transferred material takes depends strongly on the magnetic field of the white dwarf. For systems with fields in the range $B \sim 10 - 70$ MG (the AM Her stars or polars), the magnetic interaction between the stars locks the white dwarf into synchronous rotation with the binary orbit and material flows from the secondary to the primary along the field lines. For systems with low surface fields (e.g. the dwarf novae), the flow is via an accretion disc that forms around the white dwarf. For systems with intermediate-strength fields (the intermediate polars – IPs[†]), the white dwarf rotation is not synchronized with the orbit and the rotating magnetosphere of the primary is able to disrupt the disc out to a radius where the ram pressure of the accretion flow balances the magnetic pressure; sufficiently strong fields may be able to prevent the formation of an accretion disc altogether. For a detailed review of the subject, see Warner (1995).

A knowledge of the magnetic field strength in CVs is thus vitally important, not only in determining the mode of accretion, but also in studying the physical conditions within the accreting regions and the binary evolution. Unfortunately, although the polars are well-studied in this respect (see Wickramasinghe 1993), it has proved remarkably difficult to determine the field strengths in IPs. To date, only 3 of the 13 definite IPs have measured magnetic field strengths: PQ Gem ($B \sim$ 9–21 MG; Vath, Chanmugam & Frank 1996), BG CMi ($B \sim$ 4 MG; Chanmugam et al. 1990) and RXJ 1712.6–2414 ($B \sim$ 8 MG; Buckley et al. 1995). These three estimates are based upon model-dependent fits to broad-band polarimetric observations and are hence uncertain; no other IPs have been observed to show significant polarization (Cropper 1986; Stockman et al. 1992).

Clearly, a more accurate and more sensitive means of determining the white dwarf magnetic field strengths in IPs is required. One of the most successful methods of determining field strengths in polars has been to measure the cyclotron humps in their spectra (e.g. Visvanathan & Wick-ramasinghe 1979; Cropper et al. 1988; Beuermann, Thomas & Schwope 1988; Wickramasinghe, Ferrario & Bailey 1989; Ferrario et al. 1989; Cropper et al. 1990; Wickramasinghe et al. 1991; Cropper & Wickramasinghe 1993). Cyclotron radiation from electrons gyrating around magnetic field-lines occurs at discrete harmonics of the fundamental. The distribution of power is shifted to higher harmonics as the electron energy increases. At the temperatures expected in the accretion region of the white dwarf, the harmonics are broadened and merge, causing humps in the spectrum. The spacing of the humps provides tight constraints on the field strength and their breadth some indication of the temperature in the emitting region (e.g. Wickramasinghe & Meggitt 1982; Barrett & Chanmugam 1985).

---

★ Present address: Systems Engineering and Assessment Ltd., Beckington Castle, PO Box 800, Bath BA3 6TB
[†] Some authors (e.g. Patterson 1994) refer to these objects as DQ Her stars. In this paper, we will adhere to the scheme favoured by Warner (1995), where DQ Her stars are a small sub-set of IPs which are rapidly rotating and lack hard X-ray emission.

Not all polars show clear evidence for cyclotron humps in their optical spectra. With fields in the range 10–70 MG, the fundamental cyclotron frequencies in polars are in the infrared from approximately 10 to 1 microns. Since harmonics at lower frequencies are more easily resolvable than those at higher frequencies, Bailey, Ferrario & Wickramasinghe (1991) and Ferrario, Bailey & Wickramasinghe (1993) observed polars in the infrared part of the spectrum to detect cyclotron humps and successfully measured the magnetic fields of AM Her (14.5 MG) and ST LMi (12 MG) in this way. This result prompted us to observe IPs in the infrared, since if they have lower fields than polars one would expect their cyclotron humps to be more visible in the infrared, where the cyclotron harmonics are closer to their fundamental frequency and therefore more easily resolvable, than in the optical. The initial results of our study (obtained as part of an on-going survey of CVs in the infrared – see paper I; Dhillon & Marsh 1995) were promising, as we (incorrectly) appeared to have detected part of a cyclotron hump in a K-band spectrum of BG CMi (Dhillon & Marsh 1993). In this paper (paper II), we present the results of our follow-up observations of the three IPs PQ Gem, BG CMi and EX Hya. For a detailed list of system parameters and previous work on each of these objects, see table 7.1 in Warner (1995).

## 2 OBSERVATIONS

On the nights of 1993 February 7 & 8 we obtained spectra of the IPs PQ Gem, BG CMi, EX Hya and the M-dwarfs Gl203 and Gl581 with the Cooled Grating Spectrometer (CGS4) on the United Kingdom 3.8 m Infrared Telescope (UKIRT) on Mauna Kea, Hawaii. CGS4 is a 1–5 micron spectrometer containing an InSb array with 58×62 pixels. The 75 l/mm grating with the 150 mm camera gave a resolution of approximately 900 $km\,s^{-1}$. Optimum spectral sampling and bad pixel removal were obtained by mechanically shifting the array over two pixels in the dispersion direction in steps of 0.5 pixels. To cover the wavelength range 0.97–2.45 micron with overlap required 6 different grating settings, centred at 1.06 microns (second order), 1.23 microns (second order), 1.41 microns (second order), 1.59 microns (second order), 1.87 microns (first order) and 2.25 microns (first order). We employed the non-destructive readout mode of the detector in order to reduce the readout noise. The slit width was 3 arcseconds (projecting to approximately 1 pixel on the detector) and was oriented at the parallactic angle throughout the run. In order to compensate for fluctuating atmospheric $OH^-$ emission lines (Ramsay, Mountain & Geballe 1992) we took relatively short exposures and nodded the telescope primary so that the object spectrum switched between two different spatial positions on the detector. A full journal of observations is presented in table 1.

## 3 DATA REDUCTION

The initial steps in the reduction of the 2D frames were performed automatically by the CGS4 data reduction system (Daley & Beard 1994). These were: the application of the bad pixel mask, bias and dark frame subtraction, flat field division, interlacing integrations taken at different detector positions, and co-adding and subtracting nodded frames. Further details of the above procedures may be found in the review by Joyce (1992). In order to obtain 1D data, we subtracted the residual sky and then extracted the spectra by summing the counts in the three columns containing the object flux.

There were three stages to the calibration of the spectra. The first was the calibration of the wavelength scale using argon arc lamp exposures. The second-order polynomial fits to the arc lines always yielded an error of less than 0.0004 microns (rms). The next step was the removal of the ripple arising from variations in the star brightness between integrations (i.e. at different detector positions). These variations were due to changes in the seeing, sky transparency and the slight motion of the stellar image relative to the slit. We discovered that the amplitude of the ripple varied across the spectrum and we were thus forced to apply a correction for this based on a linear interpolation of the ripple profiles at either end of the spectrum. The final step in the spectral calibration was the removal of telluric atmospheric features and flux calibration. This was performed by dividing the spectra to be calibrated by the spectrum of an F-type standard, observed at a similar (typically within 0.1) airmass, with its prominent stellar features masked out. We then multiplied the result by the known flux of the standard at each wavelength, determined using a black body function set to the same temperature and magnitude as the standard. As well as providing flux calibrated spectra, this procedure also removed atmospheric absorption features from the object spectra.

## 4 RESULTS

Figure 1 shows the 0.97–2.45 micron spectra of the IPs PQ Gem, BG CMi and EX Hya together with the spectra of M3+ and M3.5 dwarf stars. We also show the spectrum of an F6V star, which indicates the location of telluric absorption features; spectral features within the strongest absorption bands are highly uncertain. In table 2 we list the wavelengths, equivalent widths and velocity widths of the most prominent spectral lines identified in figure 1.

### 4.1 Emission lines

The most prominent features in the IP spectra are the Paschen, Brackett and He I emission lines. Their large velocity widths and the fact that they are so strongly in emis-sion suggest an origin in the accretion stream and/or disc. This result is in broad agreement with optical observations of these IPs, which show Balmer and He I emission lines of similar strength from the accretion stream and/or disc (e.g. Smith et al. 1996; de Martino et al. 1995; Hellier et al. 1987). Optical observations, however, also show higher-excitation emission lines (such as He II $\lambda$4686 Å), something not visible in the infrared spectra. The infrared emission lines appear similar in all three IPs, with the exception of the greater velocity widths of the lines in EX Hya (due to its higher inclination). The spectral resolution is insufficient, however, to resolve the EX Hya emission lines into the double-peaked profiles observed in the optical (Hellier et al. 1987).

### 4.2 Absorption lines/bands

The continuum of EX Hya is dominated by strong water absorption bands around 1.4, 1.7 and 2.3 microns. These features are also observed in the M-dwarf spectra of figure 1, implying that we have detected the secondary star in EX Hya. There is also some evidence for weak Na I and $^{12}$CO absorption from the secondary, although the latter is complicated by the presence of the Pfund-limit at 2.28 microns (which might also be responsible for the scatter in the spectral-type determinations of Dhillon & Marsh 1995). Note that Smith, Collier Cameron & Tucknott (1993) also report a detection of the secondary in EX Hya in the optical from skew-mapping experiments. A comparison of the spectrum of EX Hya with the M3+ and M3.5 dwarf spectra plotted in figure 1 shows that the water band longward of 2.29 microns (an excellent indicator of spectral type; Dhillon & Marsh 1995) is not as steep in EX Hya as in the M-dwarfs. Assuming a flat accretion disc contribution in the K-band, this implies the spectral-type of the secondary is slightly earlier than M3+ but certainly later than M0 (see Dhillon & Marsh 1995). This determination is in broad agreement with what might be expected if one were to adopt main-sequence assumptions for the lobe-filling secondary in a 1.63 hour binary (see figure 2.45 in Warner 1995). By subtracting varying secondary star contributions and inspecting the residuals, we estimate the K-band contribution of the secondary star in EX Hya to be 75±25%, assuming the secondary is of the same spectral type as Gl581 (M3+V).

The infrared spectra of PQ Gem and BG CMi show no evidence for absorption features from the secondary star, although higher signal-to-noise spectra would be required to be certain. With orbital periods of 5.18 and 3.24 hours, respectively, one might have expected to see absorption lines/bands of Na I, Ca I, $^{12}$CO and H$_2$O from an early-type M-dwarf/late-type K-dwarf, as observed in the infrared spectra of dwarf novae (Dhillon & Marsh 1995). The absence of secondary star features in the infrared spectrum of PQ Gem is not in agreement with the results of Smith et al. (1996), who reported a detection of the $\lambda$8190 Å Na I doublet in the same object. It should be stressed, however, that the detection by Smith et al. (1996) is marginal. The fact that no secondary star features are visible in PQ Gem and BG CMi implies that the disc/stream must contribute a larger fraction of the infrared light than the disc/stream in EX Hya, most probably due to differing mass transfer rates. By subtracting varying secondary star contributions and inspecting the residuals, we estimate upper limits to the K-band contribution of the secondary star of ∼50% and ∼30% in BG CMi and PQ Gem, respectively, assuming the secondary is of the same spectral type as Gl581 (M3+V).

### 4.3 Continuum flux distributions

The three IPs exhibit very different continuum flux distributions. As described in section 4.2, the continuum of EX Hya is dominated by water absorption bands from the M-dwarf secondary and shows no evidence for cyclotron humps. Sherrington et al. (1980) found a simple Planckian disc (without truncating the inner regions) provides a good fit to the EX Hya flux distribution from the ultraviolet to the infrared, implying a dominant accretion disc contribution in the infrared. Our results are in conflict with this conclusion, as the secondary star appears to be a significant contributor to the infrared light in EX Hya.

In contrast to the K-band spectrum of BG CMi presented by Dhillon & Marsh (1993), the spectrum of BG CMi in figure 1 does not show cyclotron humps. This prompted us to re-examine the original data of Dhillon & Marsh (1993), where we discovered an error – the K-band spectrum presented by Dhillon & Marsh (1993) is contaminated by the light of a nearby (saturated) star, rendering the data invalid. The data presented in figure 1 does not suffer from this problem. The infrared continuum of BG CMi can be represented by a power-law of the form $f_\nu \propto \nu^{0.6}$ or by a blackbody of temperature 4500 K. The power-law shape is supported by the optical continuum presented by McHardy et al. (1987), who measured $f_\nu \propto \nu^{\sim 0.4}$. These slopes are in general agreement with what might be expected of a steady-state disc at optical-infrared wavelengths (Lynden-Bell 1969).

The infrared continuum of PQ Gem is also devoid of cyclotron humps and can be represented by a power law of the form $f_\nu \propto \nu^{0.9}$ or by a blackbody of temperature 5100 K. A blackbody continuum is supported by the optical spectrum of Smith et al. (1996), which shows a turnover in the spectrum below 1 micron consistent with a ∼5000 K blackbody. Hence the optical-infrared continuum of PQ Gem is not well-represented by a power-law (and hence not well represented by a steady-state disc).

## 5 DISCUSSION

It appears that the infrared spectra of PQ Gem and BG CMi do not show cyclotron humps. This was thought to be the case with many weak-field polars until Cropper et al. (1988) demonstrated that humps are common in polars if the emission lines and low-order continuum are subtracted from optical spectra of sufficient signal-to-noise, phase discrimination and wavelength coverage. Following Cropper et al. (1988), we therefore created a synthesized spectrum by fitting a fifth-order polynomial to the continuum and multiple Gaussians to the emission and absorption features of PQ Gem and BG CMi. The result was subtracted from each spectrum and the residual binned in the spectral direction and divided by the polynomial fit to the continuum to produce the cyclotron-hump spectra plotted as a percentage of the continuum flux in figure 2. We did not attempt this analysis with EX Hya due to the uncertain secondary star contribution.

4  *V. S. Dhillon et al.*

The spectra in figure 2 do show features which might be attributable to cyclotron humps (which would appear equally spaced in wavenumber). A comparison with the F-star spectrum in figure 1, however, reveals that the largest features are due to residual telluric features (e.g. around 1.4 and 1.9 microns in PQ Gem) and minor errors in the synthesized spectrum (e.g. around 1.1 microns in PQ Gem). Ignoring these features leaves little evidence for cyclotron emission. From the lowermost spectra in figure 2 we can constrain any cyclotron emission present to contribute less than ∼3% of the infrared continuum flux in both PQ Gem and BG CMi.

The fact that cyclotron humps are not visible in our spectra may not be surprising given the strength of circular polarization observed in BG CMi (−1.7% in the J-band and −4.2% in the H-band; Penning, Schmidt & Liebert 1986, West, Berriman & Schmidt 1987), PQ Gem (2.1% in the I-band; Piirola, Hakala & Coyne 1993, Rosen, Mittaz & Hakala 1993) and EX Hya (−0.02% in the V+R+I bands; Cropper 1986). If we make the reasonable assumption that the infrared circular polarization is due to cyclotron emission (West, Berriman & Schmidt 1987; Chanmugam et al. 1990), then in the case of BG CMi our infrared spectrum is probably of insufficient quality to detect cyclotron humps at the level indicated by the circular polarization and higher signal-to-noise observations would be required to detect them. In the case of PQ Gem, it is also possible that the cyclotron emission becomes optically thick longward of the I-band and thermalized, resulting in a blackbody continuum and no cyclotron humps. Furthermore, the circular polarization from PQ Gem is modulated at the primary's rotation period of 833 s and our observing technique of integrating at each of the 6 grating positions for one full spin cycle might have smeared any cyclotron humps present. (This effect is unlikely to account for the absence of humps in BG CMi as its circular polarization appears to be unmodulated with spin phase.) Finally, in the case of EX Hya, which exhibits virtually no circular polarization (implying a weaker magnetic field than in PQ Gem and BG CMi), the fundamental is likely to be far in the infrared and any weak harmonics present in the 0.97–2.45 micron range would be unresolvable.

We conclude, having observed two of arguably the three strongest field IPs in the optimum part of the spectrum and not having detected cyclotron humps, that measuring accurate magnetic field strengths in this manner will probably not be feasible. To maximise the chances of success, much higher signal-to-noise observations would be required, the secondary star would have to be detected so as to correct for its contribution and great care would have to be taken with telluric absorption correction. More realistically, we believe the only way to proceed is through infrared circular spectropolarimetry, which would only be sensitive to cyclotron emission and hence insensitive to the contributions of the secondary star and telluric absorption.

ACKNOWLEDGEMENTS

We would like to thank Robert Smith for his comments on a draft of this paper. UKIRT is operated by the Joint Astronomy Centre on behalf of the Particle Physics and Astronomy Research Council.ACKNOWLEDGEMENTS

We would like to thank Robert Smith for his comments on a draft of this paper. UKIRT is operated by the Joint Astronomy Centre on behalf of the Particle Physics and Astronomy Research Council.

**REFERENCES**

Bailey J. A., Ferrario L., Wickramasinghe D. T., 1991, MNRAS, 251, 37P
Barrett P. E., Chanmugam G., 1985, ApJ, 298, 743
Beuermann K., Thomas H.-C., Schwope A. D., 1988, A&A, 195, L15
Boeshaar P. C., 1976, PhD thesis, Ohio State University
Buckley D. A. H., Sekiguchi K., Motch C., O'Donoghue D., Chen A.-L., Schwarzenberg-Czerny A., Pietsch W., Harrop-Allin M., 1995, MNRAS, 275, 1028
Chanmugam G., Frank J., King A. R., Lasota J.-P., 1990, ApJ, 350, L13
Cropper M., Wickramasinghe D. T., 1993, MNRAS, 260, 696
Cropper M. et al., 1988, MNRAS, 236, 29P
Cropper M. et al., 1990, MNRAS, 245, 760
Cropper M., 1986, MNRAS, 222, 225
Daley P. N., Beard S. M., 1994, Starlink User Note 27, DRAL
de Martino D., Mouchet M., Bonnet-Bidaud J. M., Vio R., Rosen S. R., Mukai K., Augusteijn T., Garlick M. A., 1995, A&A, 298, 849
Dhillon V. S., Marsh T. R., 1993, in Regev O., Shaviv G., eds, Cataclysmic Variables and Related Physics. Inst. Phys. Publ., Bristol, p. 34
Dhillon V. S., Marsh T. R., 1995, MNRAS, 275, 89
Ferrario L., Bailey J. A., Wickramasinghe D. T., 1993, MNRAS, 262, 285
Ferrario L., Wickramasinghe D. T., Bailey J. A., Tuohy I. R., Hough J. H., 1989, ApJ, 337, 832
Friend M. T., Martin J. S., Smith R. C., Jones D. H. P., 1990, MNRAS, 246, 637
Hellier C., Mason K. O., Rosen S. R., Cordova F. A., 1987, MNRAS, 228, 463
Hoffleit D., Jaschek C., 1982, The Bright Star Catalogue. Yale University Observatory, Connecticut
Joyce R. R., 1992, in Howell S. B., ed, Astronomical CCD Observing and Reduction Techniques. ASP Conference Series, Volume 23, p. 258
Lynden-Bell D., 1969, Nat, 233, 690
McHardy I. M., Pye J. P., Fairall A. P., Menzies J. W., 1987, MNRAS, 225, 355
Patterson J., 1994, PASP, 106, 209
Penning W. R., Schmidt G. D., Liebert J., 1986, ApJ, 301, 881
Piirola V., Hakala P. J., Coyne G. V., 1993, ApJ, 410, L107
Ramsay S. K., Mountain C. M., Geballe T. R., 1992, MNRAS, 259, 751
Rosen S. R., Mittaz J. P. D., Hakala P. J., 1993, MNRAS, 264, 171
Sherrington M. R., Lawson P. A., King A. R., Jameson R. F., 1980, MNRAS, 191, 185
Smith R. C., Collier Cameron A., Tucknott D. S., 1993, in Regev O., Shaviv G., eds, Cataclysmic Variables and Related Physics. Inst. Phys. Publ., Bristol, p. 70
Smith R. C., Sarna M. J., Catalan M. S., Jones D. H. P., 1996, MNRAS, submitted
Stockman H. S., Schmidt G. D., Berriman G., Liebert J., Moore R. L., Wickramasinghe D. T., 1992, ApJ, 401, 628
Vath H., Chanmugam G., Frank J., 1996, ApJ, 457, 407
Visvanathan N., Wickramasinghe D. T., 1979, Nat, 281, 47
Warner B., 1995, Cataclysmic Variable Stars. Cambridge University Press, Cambridge
West S. C., Berriman G., Schmidt G. D., 1987, ApJ, 322, L35
Wickramasinghe D. T., Meggitt S. M. A., 1982, MNRAS, 198, 975
Wickramasinghe D. T., Ferrario L., Bailey J. A., 1989, ApJ, 342, L35
Wickramasinghe D. T., Cropper M., Mason K. O., Garlick M., 1991, MNRAS, 250, 692

Wickramasinghe D. T., 1993, in Regev O., Shaviv G., eds, Cataclysmic Variables and Related Physics. Inst. Phys. Publ., Bristol, p. 213

**Table 1.** Journal of observations.

| Object | $\lambda_{\text{central}}$ ($\mu$m) | Date | UT start | UT end |
|---|---|---|---|---|
| PQ Gem | 1.06 | 08/02/93 | 08:17 | 08:32 |
| PQ Gem | 1.23 | 08/02/93 | 09:58 | 10:14 |
| PQ Gem | 1.41 | 08/02/93 | 10:49 | 11:05 |
| PQ Gem | 1.59 | 08/02/93 | 11:10 | 11:25 |
| PQ Gem | 1.87 | 08/02/93 | 07:07 | 07:23 |
| PQ Gem | 2.25 | 08/02/93 | 09:08 | 09:24 |
| BG CMi | 1.06 | 07/02/93 | 08:58 | 09:29 |
| BG CMi | 1.06 | 08/02/93 | 08:36 | 08:56 |
| BG CMi | 1.23 | 07/02/93 | 10:36 | 11:11 |
| BG CMi | 1.23 | 08/02/93 | 10:15 | 10:35 |
| BG CMi | 1.41 | 07/02/93 | 11:25 | 11:57 |
| BG CMi | 1.59 | 07/02/93 | 12:11 | 12:43 |
| BG CMi | 1.87 | 07/02/93 | 08:03 | 08:35 |
| BG CMi | 1.87 | 08/02/93 | 07:27 | 07:58 |
| BG CMi | 2.25 | 07/02/93 | 09:43 | 10:14 |
| BG CMi | 2.25 | 08/02/93 | 09:25 | 09:45 |
| EX Hya | 1.06 | 08/02/93 | 13:05 | 13:21 |
| EX Hya | 1.23 | 08/02/93 | 13:59 | 14:16 |
| EX Hya | 1.41 | 08/02/93 | 14:29 | 14:45 |
| EX Hya | 1.59 | 08/02/93 | 14:59 | 15:16 |
| EX Hya | 1.87 | 08/02/93 | 12:35 | 12:49 |
| EX Hya | 2.25 | 08/02/93 | 13:32 | 13:48 |

Table 2. Wavelengths, equivalent widths and velocity widths of the most prominent lines in the infrared spectra of the intermediate polars PQ Gem, BG CMi and EX Hya and the dwarf stars Gl581 (M3+; Friend et al. 1990) and Gl203 (M3.5; Boeshaar 1976). The line identifications have been based upon the list presented by Dhillon & Marsh (1995) and references therein. The wavelengths given for the $^{12}$CO lines refer to the band-heads. The vertical bars following lines of similar wavelength indicate that the measurements apply to the entire blend. The two-letter codes indicate that the line is either not present (np) or that the line is present but is not measurable (nm).

| Line | $\lambda$ $\mu$m | PQ Gem EW Å | FWHM km s$^{-1}$ | BG CMi EW Å | FWHM km s$^{-1}$ | EX Hya EW Å | FWHM km s$^{-1}$ | Gl581 EW Å | FWHM km s$^{-1}$ | Gl203 EW Å | FWHM km s$^{-1}$ |
|---|---|---|---|---|---|---|---|---|---|---|---|
| Pa-$\delta$ | 1.0049 | 30±3 | 1700±200 | 15±4 | 1100±300 | 41±3 | 1900±100 | np | np | np | np |
| HeI | 1.0830 | 76±2 | 1000±100 | 78±3 | 1000±100 | 91±2 | 1600±100 | np | np | np | np |
| Pa-$\gamma$ | 1.0938 | 36±2 | 1300±100 | 29±3 | 1200±100 | 69±2 | 2000±100 | np | np | np | np |
| Pa-$\beta$ | 1.2818 | 45±1 | 1100±100 | 34±2 | 1200±100 | 68±2 | 2000±100 | np | np | np | np |
| Br-limit | 1.4584 | nm | nm | nm | nm | nm | nm | np | np | np | np |
| Br-20 | 1.5192 | nm | nm | nm | nm | nm | nm | np | np | np | np |
| Br-19 | 1.5261 | nm | nm | nm | nm | nm | nm | np | np | np | np |
| Br-18 | 1.5342 | nm | nm | nm | nm | nm | nm | np | np | np | np |
| Br-17 | 1.5439 | nm | nm | nm | nm | nm | nm | np | np | np | np |
| Br-16 | 1.5557 | nm | nm | nm | nm | nm | nm | np | np | np | np |
| Br-15 | 1.5701 | nm | nm | nm | nm | nm | nm | np | np | np | np |
| Br-14 | 1.5881 | nm | nm | nm | nm | nm | nm | np | np | np | np |
| Br-13 | 1.6109 | nm | nm | nm | nm | nm | nm | np | np | np | np |
| Br-12 | 1.6407 | nm | nm | nm | nm | nm | nm | np | np | np | np |
| Br-11 | 1.6807 | 15±5 | 1200±400 | 12±5 | 1400±600 | 31±6 | 2300±600 | np | np | np | np |
| Br-10 | 1.7362 | 23±6 | 1100±300 | 14±5 | 1100±400 | 36±6 | 2800±500 | np | np | np | np |
| Br-$\epsilon$ | 1.8174 | 26±7 | 1300±400 | 10±7 | 900±400 | 34±8 | 2900±900 | np | np | np | np |
| Pa-$\alpha$ | 1.8751 | 178±22 | 1000±100 | 92±15 | 800±100 | 117±18 | 1800±200 | np | np | np | np |
| Br-$\delta$ | 1.9446 | 54±7 | 1200±200 | 32±7 | 1000±200 | 61±9 | 2000±300 | np | np | np | np |
| HeI | 2.0587 | 24±2 | 1000±100 | 34±6 | 700±200 | 25±2 | 1600±200 | np | np | np | np |
| Br-$\gamma$ | 2.1655 | 56±3 | 1100±100 | 62±7 | 1000±100 | 72±3 | 2100±100 | np | np | np | np |
| NaI | 2.2062 | np | np | np | np | −2±1 | 1000±200 | −4±1 | 1000±100 | −3±1 | 900±100 |
| NaI | 2.2090 | | | | | | | | | | |
| CaI | 2.2614 | np | np | np | np | nm | nm | −2±1 | 900±200 | −3±1 | 1000±200 |
| CaI | 2.2631 | | | | | | | | | | |
| CaI | 2.2657 | | | | | | | | | | |
| Pf-limit | 2.2788 | nm | nm | nm | nm | nm | nm | np | np | np | np |
| $^{12}$CO | 2.2935 | np | np | np | np | −3±2 | nm | −7±2 | nm | −8±2 | nm |
| $^{12}$CO | 2.3227 | np | np | np | np | nm | nm | nm | nm | nm | nm |
| $^{12}$CO | 2.3525 | np | np | np | np | nm | nm | nm | nm | nm | nm |
| $^{12}$CO | 2.3830 | np | np | np | np | nm | nm | nm | nm | nm | nm |

Figure 1. Infrared spectra of the intermediate polars PQ Gem, BG CMi and EX Hya and the dwarf stars Gl581 (M3+; Friend et al. 1990) and Gl203 (M3.5; Boeshaar 1976). The spectra have been normalized by dividing by the flux at 2.24 microns and then offset by adding a multiple of 0.75 to each spectrum. Also shown is the spectrum of an F6V star (HR1543; Hoffleit & Jaschek 1982), normalized by dividing by a spline fit to its continuum, which indicates the location of telluric absorption features. Each spectrum consists of 6 sub-spectra which have been merged by matching the flux levels in overlapping regions (illustrated by the vertical dotted lines in the lowermost spectrum) and then cast into 500 wavelength bins.

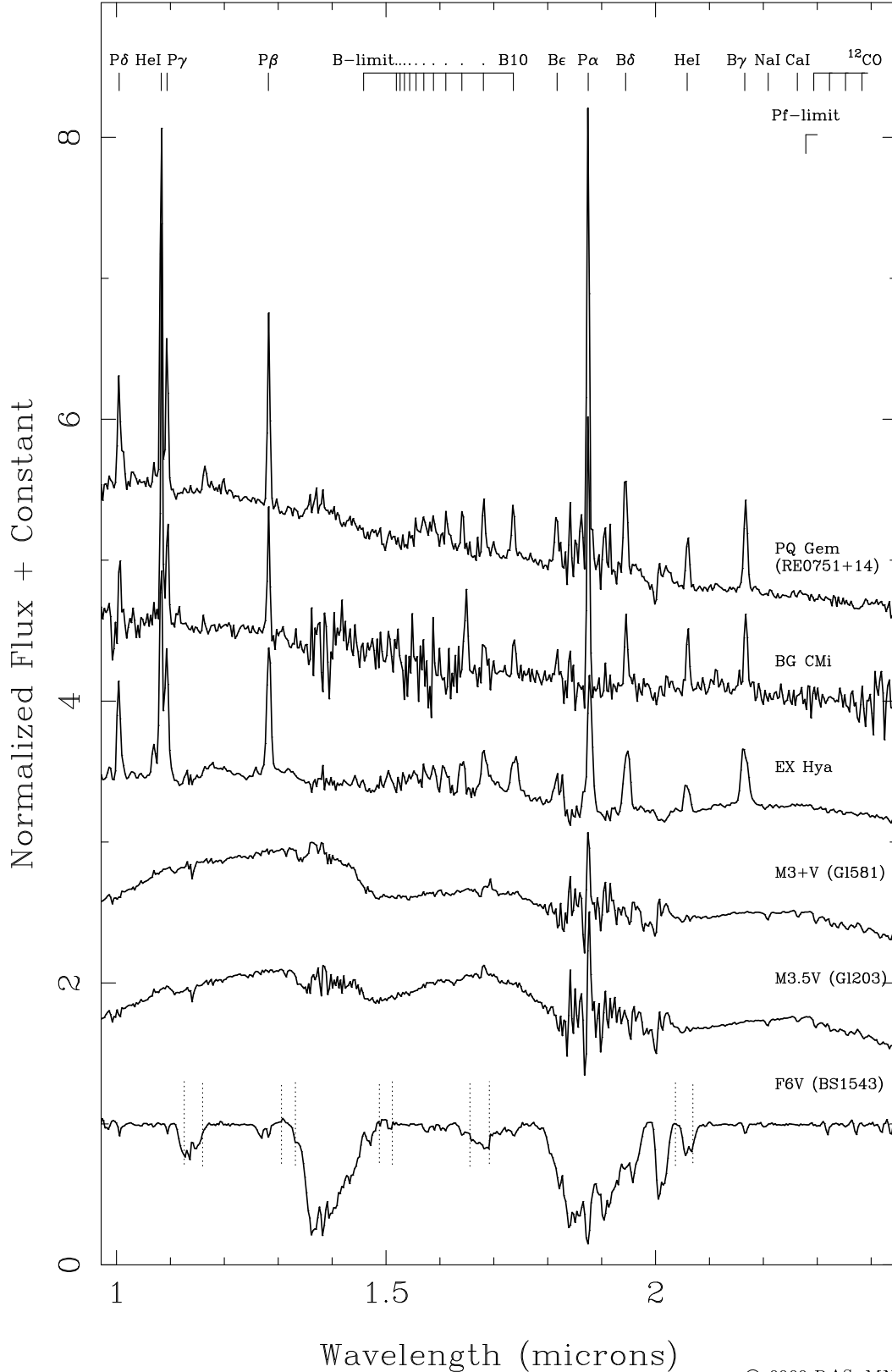

**Figure 2.** Cyclotron hump spectra of PQ Gem (left-hand panel) and BG CMi (right-hand panel), following the technique of Cropper et al. (1988). The uppermost spectrum is the 0.97–2.45 micron data of figure 1, plotted versus wavenumber (so that any cyclotron humps present would appear equally spaced). Beneath this is a synthesized spectrum, consisting of Gaussian emission/absorption lines and a fifth-order polynomial continuum. The residual spectrum, shown in the third plot, is cast into 50 wavenumber bins and divided by the polynomial fit to the continuum of the uppermost spectrum, producing the cyclotron hump spectrum plotted as a percentage of the continuum flux at the bottom. A comparison with the F-star spectrum in figure 1 reveals that the humps in the lowermost spectrum are due to residual telluric features and not cyclotron emission.

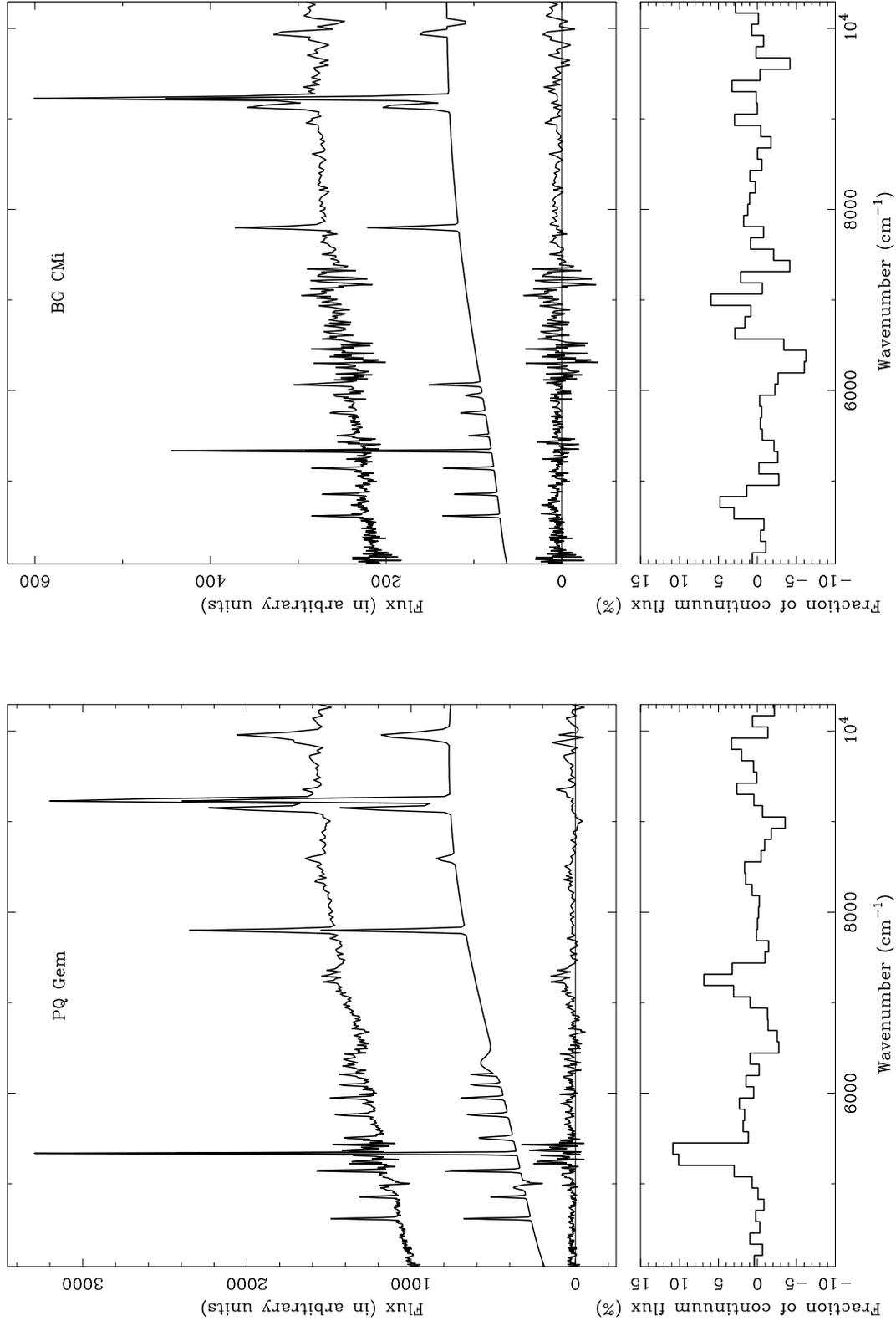